\documentclass[prb, reprint, twocolumn, showpacs, amsmath, amssymb]{revtex4-1}

\usepackage{graphicx}% Include figure files
\usepackage{dcolumn}% Align table columns on decimal point
\usepackage{bm}% bold math
\usepackage{multirow}
\usepackage{epstopdf}
\usepackage{textcomp, amsmath}
\usepackage{threeparttable}

\usepackage{color}
\usepackage{graphicx}

\begin{document}
%================= Title, affiliations etc =============================================%
%\title{Diffusion Monte Carlo Calculations of the HCP to BCC phase transition in beryllium}
\title{Beyond chemical accuracy: The pseudopotential approximation in diffusion Monte Carlo calculations of the HCP to BCC phase transition in beryllium}

\author{Luke Shulenburger}
\email{lshulen@sandia.gov}
%\affiliation{Sandia National Laboratories, Albuquerque, New Mexico 87185, USA}

\author{Thomas R. Mattsson}
\email{trmatts@sandia.gov}
%\affiliation{Sandia National Laboratories, Albuquerque, New Mexico 87185, USA}

\author{M. P. Desjarlais}
\email{mpdesja@sandia.gov}
\affiliation{Sandia National Laboratories, Albuquerque, New Mexico 87185, USA}

\date{\today}

%============================= ABSTRACT ====================================%
\begin{abstract}
Motivated by the disagreement between recent diffusion Monte Carlo calculations and experiments on the phase transition pressure between the ambient and beta-Sn phases of silicon, we present a study of the HCP to BCC phase transition in beryllium.  This lighter element provides an opportunity for directly testing many of the approximations required for calculations on silicon and may suggest a path towards increasing the practical accuracy of diffusion Monte Carlo calculations of solids in general.  We demonstrate that the single largest approximation in these calculations is the pseudopotential approximation.  After removing this we find excellent agreement with experiment for the ambient HCP phase and results similar to careful calculations using density functional theory for the phase transition pressure.
\end{abstract}

\preprint{{\large SAND2015-20744 J}}

\maketitle

%=========================== INTRODUCTION ================================%

%%{\bf Introduction}
While diffusion quantum Monte Carlo (DMC) has proven to be among the most accurate methods for electronic structure calculations,\cite{qmc-solids-prb,dmc-solids-review,ncstate-review,cambridge-review} some recent results have shown that the method may have some difficulty in predicting energetics between competing structures in condensed phase materials heavier than hydrogen.\cite{hennig-silicon,sorella-silicon,maezono-silicon}  Although the method is systematically improvable, these studies have all employed two important approximations.  The first of these is the use of pseudopotentials to remove the need to explicitly treat the chemically inert core electrons, and the second is use of a single Slater-determinant trial wavefunction to restrict the nodal surface that constrains the diffusion Monte Carlo calculations.  Given the computational expense of DMC, it is crucial for the future use of the method to understand the relative importance of these approximations and if possible understand how to reduce their magnitude.

In this paper, we address the importance of these approximations for the conceptually simple but computationally challenging solid-solid phase transition of beryllium from the ambient HCP phase to BCC at high pressure.  Beryllium is an interesting material due to its importance as a lightweight material with relevance to a number of high pressure applications, including the recently proposed MagLIF and NIF fusion concepts.\cite{maglif-concept,nif-concept}  Despite this technological significance, the phase diagram of beryllium at high temperatures and pressures is not tightly constrained from experiments despite a great deal of effort.\cite{benedict-be,lazicki-be,zhang-dac,evans-dac,brown-be-dynamic}  The material has also attracted numerous attempts to calculate its phase diagram using electronic structure techniques.\cite{french-be-calc,robert-be-calc,luo-alfe-be-calc,cheng-be-calc,guo-be-calc}  In addition, the phase transition from HCP to BCC structures is extraordinarily sensitive to any errors that may be present in the underlying electronic structure calculations.  

\begin{figure}
\includegraphics[width=2.3in,angle=-90]{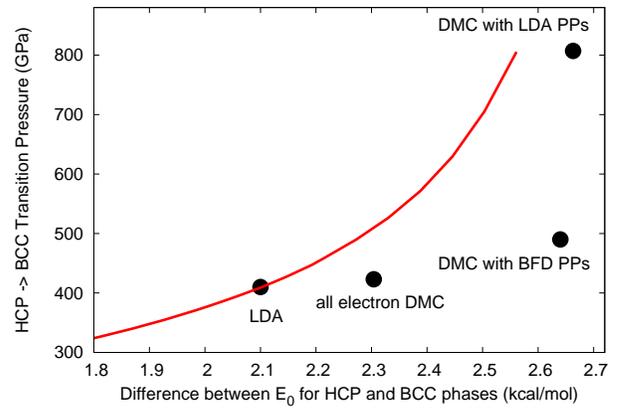}
\caption{(color online) Calculated zero temperature phase transition pressure from HCP to BCC.  Calculations were performed using the LDA functional and include a quasiharmonic treatment of the zero point motion of the beryllium.  The dot labeled LDA denotes the calculated energy difference between the zero pressure volumes of BCC and HCP.  The line shows the change in transition pressure as this difference is artifically changed.  The values of the energy difference between the equilibrium density HCP and BCC phases and corresponding transition pressures are also plotted for DMC with different choices for treating the core electrons.
\label{be-sensitivity}}
\end{figure}

To illustrate this, we present a series of density functional theory (DFT) based calculations of the zero temperature phase transition pressure.  The DFT framework requires an approximation in the choice of a functional to treat the effects of electron correlation.  Although it is possible to classify the complexity of approximations to this exchange correlation functional,\cite{jacobs-ladder} increasing this complexity does not guarantee increased accuracy.  In fact, there is no {\it a priori} way to select the most accurate functional for a specific calculation and the spread in predictions of the total energy in a solid can often be larger than the 1 kcal/mol ($\approx 43$ meV or 0.8 mHa / atom) target that is typically deemed chemical accuracy.  Here both the HCP and BCC phases are treated within the local density approximation, utilizing the quasiharmonic approximation to account for the zero-point motion of the relatively light nuclei.  This initial calculation results in a phase transition pressure of $\approx 410$ GPa, shown figure \ref{be-sensitivity} as the point labeled LDA.  However, we also consider the simplest possible error between the phases: a rigid shift of the energies of one phase compared to another, shown by the line in the same figure.  This produces an enormous change in the phase transition pressure.  In fact, as this energy difference is shifted by 0.3 kcal/mol (less than 1/3 of the value typically desired for chemical accuracy) the phase transition pressure changes by over 150 GPa.

This extreme sensitivity motivates application of diffusion Monte Carlo (DMC) to the problem.  While the size of the errors in DMC calculations of real materials has yet to be seen, the approximation made in DMC is of an entirely different class (being primarily a topological constraint on the many body wavefunction) than the one made in density functional theory.  For this reason one may seek by DMC to build a diversity of approximations, easing the uncertainty in predictions caused by reliance on a single computational technique.

As a starting point for these DMC calculations, we utilize the methodology presented in our recent paper on applying quantum Monte Carlo to solids.\cite{qmc-solids-prb}  Summarizing this briefly, we use the {\sc qmcpack} code to perform DMC calculations on periodic systems.\cite{qmcpack,gpu-qmc}  Pseudopotentials are generated using the {\sc opium} pseudopotential generation program\cite{opium} so as to accurately reproduce the results of all electron LAPW calculations for the lattice constant and bulk modulus of the ambient phase of the material.  
%Specifically, a Troullier Martins\cite{troullier-martins} construction was used with cutoff radii of 1.3 bohr for the ``s'' channel and 1.15 bohr for the ``p'' channel.  
For DMC, the pseudopotential was evaluated using a variational technique.\cite{casula-t-moves}  One and two-body Jastrow factors were used to construct a Slater-Jastrow wavefunction where the single particle orbitals were extracted from DFT calculations using the local density approximation.  The timestep and the number of supercell twists were chosen so that the total energy of the calculations was converged to within 1 meV per atom. To accurately calculate a phase transition pressure, the two body finite size effects need to be converged absolutely rather than relatively within each phase.  In the case of HCP beryllium, this required 132 atom supercells with 2-body finite size errors estimated using a combination of the model periodic Coulomb interaction\cite{MPC} and a kinetic energy correction estimated from the long range piece of the variational Monte Carlo optimized 2-body Jastrow factor.\cite{Chiesa-finite-size}  
%In this way we have systematically controlled all of the errors due to the timestep, finite size effects and wavefunction representation.  
Any remaining error in the calculations will be due either to the fixed node approximation or to the construction and evaluation of the pseudopotential.

\begin{figure}
\includegraphics[width=2.3in,angle=-90]{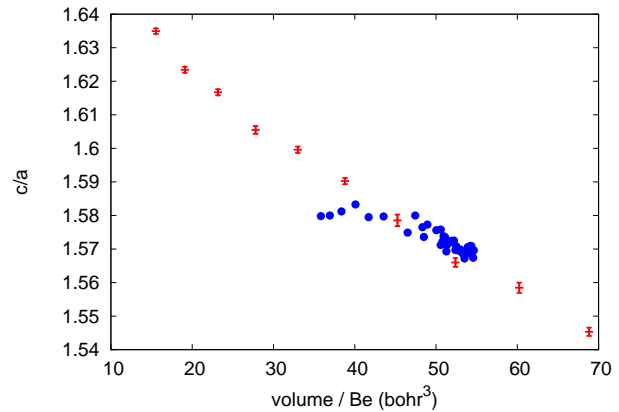}
\caption{(color online) Calculated HCP c/a from quadratic fits to the DMC energy vs volume conserving strain is shown in red with error bars.  Experimental c/a versus specific volume taken from diamond anvil cell experiments reported in reference~\onlinecite{lazicki-be} is plotted as blue circles.}
\label{covera-figure}
\end{figure}

The c/a ratio of HCP beryllium must be optimized for each volume.  This was done optimizing the DMC energy over volume conserving strains, yielding c/a ratios shown in Fig.\ref{covera-figure}.  The calculated c/a ratio at ambient pressure is 1.569 +/- 0.004, which is in excellent agreement with the experimentally measured 1.568.\cite{expt-c-a}

We now calculate the energy as a function of volume of HCP beryllium for the optimized structures, fitting the sum of resulting values and the quasiharmonic zero point energy determined within the LDA to a Vinet equation of state.\cite{vinet}  We find an equilibrium volume that is slightly too small at 52.27 $\pm$ 0.02 bohr$^3$/atom compared to the experimental 54.776 bohr$^3$/atom.  The bulk modulus is somewhat l too large at 124.21 $\pm$ 0.37 GPa, compared to the experimental value of 116.8 GPa.  Despite this disappointing performance, we continue calculating the same energy versus volume curve for the high pressure BCC structure.  This is also fit with a Vinet equation of state.  These fits are necessary to determine the phase transition pressure because our current technique does not calculate the pressure independently.  Utilizing these fits, we determine the enthalpy to be equal in the HCP and BCC phases at 807 GPa, a pressure far in excess of that calculated using density functional theory techniques.

Given the incredible sensitivity of this phase transition on any errors in the calculations, it is prudent to investigate these errors directly.  The two remaining significant sources of error are the fixed node approximation for the wavefunction and the pseudopotential error for the Hamiltonian.  In the case of solid beryllium, we believe that the nodal surface should be relatively simple and so delay this investigation.  The pseudopotential approximation is the more likely source of error and we investigate its impact by performing calculations with a bare Coulomb potential for the beryllium ions.  In this sense we are following in the footsteps of Esler {\it et al.}\cite{esler-cBN} who showed that by removing the pseudopotential approximation, highly accurate results could be obtained for the equation of state of cubic boron nitride.

There is a key difference between our approach and that taken by Esler {\it et al}.  That paper used a DFT technique, LAPW, to generate trial wavefunctions that natively satisfy the cusp condition and thereby avoid a divergence of the DMC local energy near each nucleus.  For reasons of computational efficiency, we instead have created a very hard pseudopotential for beryllium that has all 4 electrons in the valence.  This pseudopotential has only a single projector and a very small cutoff radius, resulting in a potential that is very similar to the coulomb potential further than 0.4 bohr from the nucleus.  There will naturally be a difference between the resulting wavefunctions produced in the plane wave basis set used by {\it quantum espresso}\cite{quantum-espresso} and those from an all electron code, however we can mask this by using the freedom afforded by the one body Jastrow factor to reintroduce the cusp in the wavefunction at each nucleus.  Finally, for the QMC calculations, we replace this pseudopotential with a -4/r Coulomb potential so that the resulting calculations will be entirely free of the pseudopotential approximation.

This all electron methodology poses two dangers.  Firstly, if the trial wavefunction is too poor, the timestep error will be incredibly difficult to converge and secondly, the fixed node error may be significantly increased by including the two core electrons in the valence and having to generate a nodal surface including them.  In practice, neither of these concerns appear to be a problem.  Using a timestep of 0.005 a.u. (a factor of two smaller than in the previous calculations) we were able to converge the energy to within 5 meV per atom.  This is a factor of five larger than the pseudopotential calculations, but the error was consistent across all densities and phases, thus enabling accurate calculations of the phase transition.  Concerns about the nodal surface are also unfounded.  Analyzing the wavefunctions within DFT, we expect that the core electrons will not be greatly perturbed by the pressures explored in this study and therefore this error should be relatively constant across all of our calculations.  Also, the percentage of the total energy recovered by the DMC compared to the VMC is small.  In fact it is on the order of 0.2\% in the all electron calculations compared to 0.5\% in the pseudopotential calculations.  While strictly speaking this is a measure of the quality of the trial wavefunction globally, this is in practice highly correlated with the quality of the nodal surface.

The final concern in performing these all electron calculations is their significantly greater computational cost.  In practice, the computational cost of the stochastic process grows only linearly with the number of electrons in this regime, suggesting that the dominant computation is the evaluation of the single particle orbitals.  However, the variance of the all electron calculations is a factor of 10 larger than the pseudopotential calculations, rendering the cost of obtaining a given statistical error 200 times larger than for the pseudopotential calculations.  Due to this limitation, we were unable to perform calculations on as large of a supercell, potentially resulting in larger two body finite size effects.  Following Esler et al.\cite{esler-cBN}, we utilize the information on the finite size scaling from the pseudopotential calculations to minimize these finite size effects, a prescription tested by performing all electron calculations for the largest supercell size for one density in both the HCP and BCC phases.  Specifically we use the following formula
\begin{equation}
E_{ae}^{\infty} \approx (E_{ps}^{132} - E_{ps}^{64}) + E_{ae}^{64}
\end{equation}
for the HCP phase where the subscripts ae and ps refer to all electron and pseudopotential calculations respectively and the superscripts refer to the number of atoms in those supercells.  The term in parenthesis is the estimate of the residual finite size error due to performing the all electron calculations with smaller supercells.  For the BCC phase, the all electron calculations utilized 54 atom supercells while the pseudopotential ones used 128.

\begin{figure}
\includegraphics[width=2.3in,angle=-90]{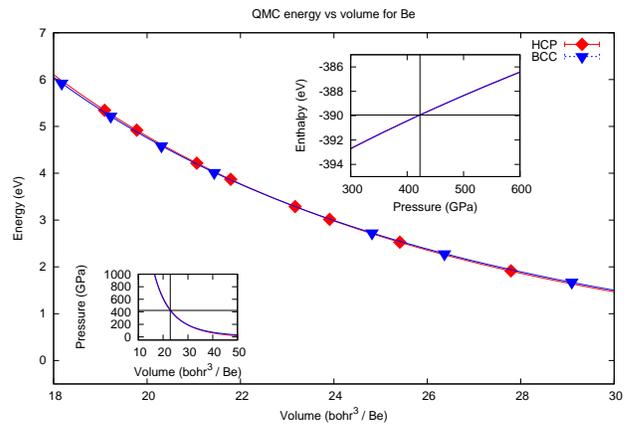}
\caption{(color online) Energy versus volume of BCC and HCP beryllium, including quasiharmonic phonon zero point energy.  Insets show the pressure vs volume and enthalpy vs pressure.  The fact that the fitting lines are on top of one another is a consequence of the nearly degenerate enthalpy of the phases.}
\label{phase-transition}
\end{figure}

Calculating the HCP energy vs volume curve, we find a marked improvement in accuracy.  The equilibrium volume is 54.87 $\pm$ 0.04 bohr$^3$, which is very close to the experimental 54.776.  The bulk modulus is similarly improved at 115.7 $\pm$ 1.0 GPa, in excellent agreement with the experimental 116.8 GPa.  Continuing the calculation to the BCC phase, we find the phase transition pressure is 423 GPa, nearly a factor of two smaller than the pseudopotential calculation and in relatively good agreement with DFT calculations.  These results are shown in Fig~\ref{phase-transition}.

The improvement in accuracy for the ambient properties of the HCP phase supports the conjecture that the predominant error in these calculations was due to the use of these particular LDA derived pseudopotentials.  The pertinent question is how to avoid this loss of accuracy in future calculations.  The idea of turning to all electron calculations is appealing, but the $Z^{5.5-6.5}$ scaling of the computational cost of all electron calculations\cite{ceperley1986statistical,ma2005all} and the vastly increased memory required to store the all electron wavefunctions render this approach untenable for all but the lightest of elements.  Clearly, better pseudopotentials are required.  Several previous works have developed pseudopotentials using a different schemes.  Predominantly there are the Hartree-Fock derived pseudopotentials of Trail and Needs,\cite{Trail-Needs-PPs}  Burkatzki, Filippi and Dolg (BFD) \cite{burkatzki-filippi-dolg-pps} and many body correlated electron pseudopotentials of Trail and Needs.\cite{Trail-Needs-CEPPs}

\begin{table*}[]
\begin{threeparttable}
%renewcommand(\arraystretch}{0.8}
\begin{tabular}{@{}ccccc@{}}
\hline
 & LDA PPs & BFD PPs &  all electron & experiment \\
\hline\hline
HCP V$_0$ (bohr$^3$) & $52.27 \pm 0.02$ & $55.19 \pm 0.01$ & $54.87 \pm 0.04$ & 54.776 \\
HCP B$_0$ (GPa) & $124.21 \pm 0.74$ & $112.99 \pm 0.43$ & $115.69 \pm 1.04$ & 116.8 \\
HCP ambient C/A & $1.569 \pm 0.004$ & $1.569 \pm 0.004$ & $1.569 \pm 0.004$ & 1.568 \\
HCP B$^\prime$ & $3.47 \pm 0.03$ & $3.39 \pm 0.02$ & $3.41 \pm 0.06$ & - \\
BCC V$_0$ (bohr$^3$) & $51.81 \pm 0.03$ & $54.60 \pm 0.01$ & $54.04 \pm 0.07$ & - \\
BCC B$_0$ (GPa) & $124.33 \pm 0.21$ & $111.83 \pm 0.21$ & $117.61 \pm 0.66$ & - \\
BCC B$^\prime$ & $3.50 \pm 0.02$ & $3.56 \pm 0.01$ & $3.60 \pm 0.05$ & - \\
Transition Pressure (GPa) & 807 & 490 & 423 & - \\
$\Delta E$ (meV / atom) & $115.5 \pm 1.0$ & $114.5 \pm 0.7$ & $99.9 \pm 2.7$ & - \\
\hline
\end{tabular}
\caption{Parameters for HCP and BCC beryllium calculated with LDA pseudopotentials, BFD pseudopotentials and without pseudopotentials.  Also listed are the calculated phase transition pressures and the change in energy per atom from HCP at its equilibrium volume to BCC at its equilibrium volume.} 
\label{results-table}
\end{threeparttable}
\end{table*}

Here we have elected to test the BFD pseudopotential because it is derived from an entirely different philosophy than the DFT one, that is it is based on reproducing total energies rather than the shape of the wavefunction for the atom.  Using this potential for the current problem did require some additional effort as these potentials are presented only in semi-local form.  Since we used {\sc quantum espresso} to generate the trial wavefunction and it uses the vastly more computationally efficient Kleinman-Bylander\cite{kleinmann-bylander} form, we were obliged to generate projectors for each of the angular momentum channels.  While this proved straightforward for Be, this is in principle a delicate operation as there exists a strong possibility of introducing ghost states for heavier elements.  The other primary complication is that the cutoff radius for the beryllium potential (determined as the radius at which the potential for all angular momentum channels is within $10^{-5}$ Ha of the bare coulomb potential) is very large, 2.28 bohr.  This means that for all but the lowest densities considered, the cores will overlap.  While this would be considered a fatal flaw for the DFT style pseudopotentials, calculations with BFD pseudopotentials often have some small overlap of the cores, while still performing well.  For this reason, we ignore the core overlap and proceed with the calculations as implemented.

Calculating the energy versus volume relation for BCC and HCP using the BFD pseudopotential produces results in much better agreement with the all electron calculations as can be seen in Table \ref{results-table}.  This is encouraging as it shows that it is possible to produce more accurate pseudopotentials for use in quantum Monte Carlo calculations of beryllium.  However, we stop short of recommending BFD pseudopotentials for all DMC calculations.  Three potential issues with the current methodology for applying DMC to condensed phases remain to be resolved.  The first is that almost all modern plane wave DFT codes require the use of Kleinman-Bylander form, raising the issue of calculating accurate projectors for the DFT calculations needed to generate trial wavefunctions.  The second issue is due to the potential incompatibility between the treatment of core electrons implied by these potentials and the density functional calculations that are generally used to generate the trial wavefunctions in these systems.  This raises the possibility of a gross incompatibility between the trial wavefunctions that would used in the DMC calculation and the pseudopotentials.  In principle this can be resolved by optimizing all parts of the trial wavefunction within variational or diffusion Monte Carlo, however this is not currently practical for calculations of condensed matter.  The final stumbling block is the large cutoff radii of the BFD pseudopotentials.  While ignoring this overlap produced reasonable results in practice for beryllium, whether this will be more generally true should be further investigated.

In conclusion, we have presented highly accurate all electron DMC calculations of the HCP to BCC phase transition for beryllium, a property requiring relative errors to be controlled to well beyond chemical accuracy.  These calculations found excellent agreement with the experimentally measured properties of the HCP phase.  Although diamond anvil cell measurements have been unable to detect this transition to pressures of 190 GPa, our calculated value is in much closer agreement with that obtained using density functional theory.  We also found that the primary error in calculating the phase transition pressure using the methodology of Shulenburger and Mattsson\cite{qmc-solids-prb} is due to the pseudopotentials used.  As this methodology is common in the QMC literature, this result is of crucial importance to the field.    Furthermore, we showed that these errors were greatly reduced when using BFD pseudopotentials and suggested several properties to be addressed in making wide use of pseudopotentials generated in this way for calculations of condensed matter.

%============================ ACKNOWLEDGEMENTS =========================%
We thank Jeongnim Kim, Andrew Baczewski, Miguel Morales and Paul Kent for stimulating discussions.  The work was supported by the NNSA Science Campaigns and LS was supported through the Predictive Theory and Modeling for Materials and Chemical Science program by the Basic Energy Sciences (BES), Department of Energy (DOE).
Sandia National Laboratories is a multiprogram laboratory managed and operated by Sandia Corporation, a wholly owned subsiderary of Lockheed Martin Corporation, for the U.S. Department of Energy's National Nuclear Security Administration under Contract No. DE-AC04-94AL85000.

% ========================     BIBLIOGRAPHY     ==========================================%
%
\bibliographystyle{apsrev4-1}

\bibliography{manuscript}

\begin{thebibliography}{37}%
\makeatletter
\providecommand \@ifxundefined [1]{%
 \@ifx{#1\undefined}
}%
\providecommand \@ifnum [1]{%
 \ifnum #1\expandafter \@firstoftwo
 \else \expandafter \@secondoftwo
 \fi
}%
\providecommand \@ifx [1]{%
 \ifx #1\expandafter \@firstoftwo
 \else \expandafter \@secondoftwo
 \fi
}%
\providecommand \natexlab [1]{#1}%
\providecommand \enquote  [1]{``#1''}%
\providecommand \bibnamefont  [1]{#1}%
\providecommand \bibfnamefont [1]{#1}%
\providecommand \citenamefont [1]{#1}%
\providecommand \href@noop [0]{\@secondoftwo}%
\providecommand \href [0]{\begingroup \@sanitize@url \@href}%
\providecommand \@href[1]{\@@startlink{#1}\@@href}%
\providecommand \@@href[1]{\endgroup#1\@@endlink}%
\providecommand \@sanitize@url [0]{\catcode `\\12\catcode `\$12\catcode
  `\&12\catcode `\#12\catcode `\^12\catcode `\_12\catcode `\%12\relax}%
\providecommand \@@startlink[1]{}%
\providecommand \@@endlink[0]{}%
\providecommand \url  [0]{\begingroup\@sanitize@url \@url }%
\providecommand \@url [1]{\endgroup\@href {#1}{\urlprefix }}%
\providecommand \urlprefix  [0]{URL }%
\providecommand \Eprint [0]{\href }%
\providecommand \doibase [0]{http://dx.doi.org/}%
\providecommand \selectlanguage [0]{\@gobble}%
\providecommand \bibinfo  [0]{\@secondoftwo}%
\providecommand \bibfield  [0]{\@secondoftwo}%
\providecommand \translation [1]{[#1]}%
\providecommand \BibitemOpen [0]{}%
\providecommand \bibitemStop [0]{}%
\providecommand \bibitemNoStop [0]{.\EOS\space}%
\providecommand \EOS [0]{\spacefactor3000\relax}%
\providecommand \BibitemShut  [1]{\csname bibitem#1\endcsname}%
\let\auto@bib@innerbib\@empty
%</preamble>
\bibitem [{\citenamefont {Shulenburger}\ and\ \citenamefont
  {Mattsson}(2013)}]{qmc-solids-prb}%
  \BibitemOpen
  \bibfield  {author} {\bibinfo {author} {\bibfnamefont {L.}~\bibnamefont
  {Shulenburger}}\ and\ \bibinfo {author} {\bibfnamefont {T.~R.}\ \bibnamefont
  {Mattsson}},\ }\href@noop {} {\bibfield  {journal} {\bibinfo  {journal}
  {Physical Review B}\ }\textbf {\bibinfo {volume} {88}},\ \bibinfo {pages}
  {245117} (\bibinfo {year} {2013})}\BibitemShut {NoStop}%
\bibitem [{\citenamefont {Foulkes}\ \emph {et~al.}(2001)\citenamefont
  {Foulkes}, \citenamefont {Mitas}, \citenamefont {Needs},\ and\ \citenamefont
  {Rajagopal}}]{dmc-solids-review}%
  \BibitemOpen
  \bibfield  {author} {\bibinfo {author} {\bibfnamefont {W.~M.~C.}\
  \bibnamefont {Foulkes}}, \bibinfo {author} {\bibfnamefont {L.}~\bibnamefont
  {Mitas}}, \bibinfo {author} {\bibfnamefont {R.~J.}\ \bibnamefont {Needs}}, \
  and\ \bibinfo {author} {\bibfnamefont {G.}~\bibnamefont {Rajagopal}},\ }\href
  {\doibase 10.1103/RevModPhys.73.33} {\bibfield  {journal} {\bibinfo
  {journal} {Rev. Mod. Phys.}\ }\textbf {\bibinfo {volume} {73}},\ \bibinfo
  {pages} {33} (\bibinfo {year} {2001})}\BibitemShut {NoStop}%
\bibitem [{\citenamefont {Koloren{\v{c}}}\ and\ \citenamefont
  {Mitas}(2011)}]{ncstate-review}%
  \BibitemOpen
  \bibfield  {author} {\bibinfo {author} {\bibfnamefont {J.}~\bibnamefont
  {Koloren{\v{c}}}}\ and\ \bibinfo {author} {\bibfnamefont {L.}~\bibnamefont
  {Mitas}},\ }\href@noop {} {\bibfield  {journal} {\bibinfo  {journal} {Reports
  on Progress in Physics}\ }\textbf {\bibinfo {volume} {74}},\ \bibinfo {pages}
  {026502} (\bibinfo {year} {2011})}\BibitemShut {NoStop}%
\bibitem [{\citenamefont {Needs}\ \emph {et~al.}(2010)\citenamefont {Needs},
  \citenamefont {Towler}, \citenamefont {Drummond},\ and\ \citenamefont
  {R{\'\i}os}}]{cambridge-review}%
  \BibitemOpen
  \bibfield  {author} {\bibinfo {author} {\bibfnamefont {R.}~\bibnamefont
  {Needs}}, \bibinfo {author} {\bibfnamefont {M.}~\bibnamefont {Towler}},
  \bibinfo {author} {\bibfnamefont {N.}~\bibnamefont {Drummond}}, \ and\
  \bibinfo {author} {\bibfnamefont {P.~L.}\ \bibnamefont {R{\'\i}os}},\
  }\href@noop {} {\bibfield  {journal} {\bibinfo  {journal} {Journal of
  Physics: Condensed Matter}\ }\textbf {\bibinfo {volume} {22}},\ \bibinfo
  {pages} {023201} (\bibinfo {year} {2010})}\BibitemShut {NoStop}%
\bibitem [{\citenamefont {Hennig}\ \emph {et~al.}(2010)\citenamefont {Hennig},
  \citenamefont {Wadehra}, \citenamefont {Driver}, \citenamefont {Parker},
  \citenamefont {Umrigar},\ and\ \citenamefont {Wilkins}}]{hennig-silicon}%
  \BibitemOpen
  \bibfield  {author} {\bibinfo {author} {\bibfnamefont {R.~G.}\ \bibnamefont
  {Hennig}}, \bibinfo {author} {\bibfnamefont {A.}~\bibnamefont {Wadehra}},
  \bibinfo {author} {\bibfnamefont {K.~P.}\ \bibnamefont {Driver}}, \bibinfo
  {author} {\bibfnamefont {W.~D.}\ \bibnamefont {Parker}}, \bibinfo {author}
  {\bibfnamefont {C.~J.}\ \bibnamefont {Umrigar}}, \ and\ \bibinfo {author}
  {\bibfnamefont {J.~W.}\ \bibnamefont {Wilkins}},\ }\href {\doibase
  10.1103/PhysRevB.82.014101} {\bibfield  {journal} {\bibinfo  {journal} {Phys.
  Rev. B}\ }\textbf {\bibinfo {volume} {82}},\ \bibinfo {pages} {014101}
  (\bibinfo {year} {2010})}\BibitemShut {NoStop}%
\bibitem [{\citenamefont {Sorella}\ \emph {et~al.}(2011)\citenamefont
  {Sorella}, \citenamefont {Casula}, \citenamefont {Spanu},\ and\ \citenamefont
  {Dal~Corso}}]{sorella-silicon}%
  \BibitemOpen
  \bibfield  {author} {\bibinfo {author} {\bibfnamefont {S.}~\bibnamefont
  {Sorella}}, \bibinfo {author} {\bibfnamefont {M.}~\bibnamefont {Casula}},
  \bibinfo {author} {\bibfnamefont {L.}~\bibnamefont {Spanu}}, \ and\ \bibinfo
  {author} {\bibfnamefont {A.}~\bibnamefont {Dal~Corso}},\ }\href@noop {}
  {\bibfield  {journal} {\bibinfo  {journal} {Phys. Rev. B}\ }\textbf {\bibinfo
  {volume} {83}},\ \bibinfo {pages} {075119} (\bibinfo {year}
  {2011})}\BibitemShut {NoStop}%
\bibitem [{\citenamefont {Maezono}\ \emph {et~al.}(2010)\citenamefont
  {Maezono}, \citenamefont {Drummond}, \citenamefont {Ma},\ and\ \citenamefont
  {Needs}}]{maezono-silicon}%
  \BibitemOpen
  \bibfield  {author} {\bibinfo {author} {\bibfnamefont {R.}~\bibnamefont
  {Maezono}}, \bibinfo {author} {\bibfnamefont {N.~D.}\ \bibnamefont
  {Drummond}}, \bibinfo {author} {\bibfnamefont {A.}~\bibnamefont {Ma}}, \ and\
  \bibinfo {author} {\bibfnamefont {R.~J.}\ \bibnamefont {Needs}},\ }\href@noop
  {} {\bibfield  {journal} {\bibinfo  {journal} {Phys. Rev. B}\ }\textbf
  {\bibinfo {volume} {82}},\ \bibinfo {pages} {184108} (\bibinfo {year}
  {2010})}\BibitemShut {NoStop}%
\bibitem [{\citenamefont {Slutz}\ and\ \citenamefont
  {Vesey}(2012)}]{maglif-concept}%
  \BibitemOpen
  \bibfield  {author} {\bibinfo {author} {\bibfnamefont {S.~A.}\ \bibnamefont
  {Slutz}}\ and\ \bibinfo {author} {\bibfnamefont {R.~A.}\ \bibnamefont
  {Vesey}},\ }\href {\doibase 10.1103/PhysRevLett.108.025003} {\bibfield
  {journal} {\bibinfo  {journal} {Phys. Rev. Lett.}\ }\textbf {\bibinfo
  {volume} {108}},\ \bibinfo {pages} {025003} (\bibinfo {year}
  {2012})}\BibitemShut {NoStop}%
\bibitem [{\citenamefont {Haan}\ \emph {et~al.}(2011)\citenamefont {Haan},
  \citenamefont {Lindl}, \citenamefont {Callahan}, \citenamefont {Clark},
  \citenamefont {Salmonson}, \citenamefont {Hammel}, \citenamefont {Atherton},
  \citenamefont {Cook}, \citenamefont {Edwards}, \citenamefont {Glenzer},
  \citenamefont {Hamza}, \citenamefont {Hatchett}, \citenamefont {Herrmann},
  \citenamefont {Hinkel}, \citenamefont {Ho}, \citenamefont {Huang},
  \citenamefont {Jones}, \citenamefont {Kline}, \citenamefont {Kyrala},
  \citenamefont {Landen}, \citenamefont {MacGowan}, \citenamefont {Marinak},
  \citenamefont {Meyerhofer}, \citenamefont {Milovich}, \citenamefont {Moreno},
  \citenamefont {Moses}, \citenamefont {Munro}, \citenamefont {Nikroo},
  \citenamefont {Olson}, \citenamefont {Peterson}, \citenamefont {Pollaine},
  \citenamefont {Ralph}, \citenamefont {Robey}, \citenamefont {Spears},
  \citenamefont {Springer}, \citenamefont {Suter}, \citenamefont {Thomas},
  \citenamefont {Town}, \citenamefont {Vesey}, \citenamefont {Weber},
  \citenamefont {Wilkens},\ and\ \citenamefont {Wilson}}]{nif-concept}%
  \BibitemOpen
  \bibfield  {author} {\bibinfo {author} {\bibfnamefont {S.~W.}\ \bibnamefont
  {Haan}}, \bibinfo {author} {\bibfnamefont {J.~D.}\ \bibnamefont {Lindl}},
  \bibinfo {author} {\bibfnamefont {D.~A.}\ \bibnamefont {Callahan}}, \bibinfo
  {author} {\bibfnamefont {D.~S.}\ \bibnamefont {Clark}}, \bibinfo {author}
  {\bibfnamefont {J.~D.}\ \bibnamefont {Salmonson}}, \bibinfo {author}
  {\bibfnamefont {B.~A.}\ \bibnamefont {Hammel}}, \bibinfo {author}
  {\bibfnamefont {L.~J.}\ \bibnamefont {Atherton}}, \bibinfo {author}
  {\bibfnamefont {R.~C.}\ \bibnamefont {Cook}}, \bibinfo {author}
  {\bibfnamefont {M.~J.}\ \bibnamefont {Edwards}}, \bibinfo {author}
  {\bibfnamefont {S.}~\bibnamefont {Glenzer}}, \bibinfo {author} {\bibfnamefont
  {A.~V.}\ \bibnamefont {Hamza}}, \bibinfo {author} {\bibfnamefont {S.~P.}\
  \bibnamefont {Hatchett}}, \bibinfo {author} {\bibfnamefont {M.~C.}\
  \bibnamefont {Herrmann}}, \bibinfo {author} {\bibfnamefont {D.~E.}\
  \bibnamefont {Hinkel}}, \bibinfo {author} {\bibfnamefont {D.~D.}\
  \bibnamefont {Ho}}, \bibinfo {author} {\bibfnamefont {H.}~\bibnamefont
  {Huang}}, \bibinfo {author} {\bibfnamefont {O.~S.}\ \bibnamefont {Jones}},
  \bibinfo {author} {\bibfnamefont {J.}~\bibnamefont {Kline}}, \bibinfo
  {author} {\bibfnamefont {G.}~\bibnamefont {Kyrala}}, \bibinfo {author}
  {\bibfnamefont {O.~L.}\ \bibnamefont {Landen}}, \bibinfo {author}
  {\bibfnamefont {B.~J.}\ \bibnamefont {MacGowan}}, \bibinfo {author}
  {\bibfnamefont {M.~M.}\ \bibnamefont {Marinak}}, \bibinfo {author}
  {\bibfnamefont {D.~D.}\ \bibnamefont {Meyerhofer}}, \bibinfo {author}
  {\bibfnamefont {J.~L.}\ \bibnamefont {Milovich}}, \bibinfo {author}
  {\bibfnamefont {K.~A.}\ \bibnamefont {Moreno}}, \bibinfo {author}
  {\bibfnamefont {E.~I.}\ \bibnamefont {Moses}}, \bibinfo {author}
  {\bibfnamefont {D.~H.}\ \bibnamefont {Munro}}, \bibinfo {author}
  {\bibfnamefont {A.}~\bibnamefont {Nikroo}}, \bibinfo {author} {\bibfnamefont
  {R.~E.}\ \bibnamefont {Olson}}, \bibinfo {author} {\bibfnamefont
  {K.}~\bibnamefont {Peterson}}, \bibinfo {author} {\bibfnamefont {S.~M.}\
  \bibnamefont {Pollaine}}, \bibinfo {author} {\bibfnamefont {J.~E.}\
  \bibnamefont {Ralph}}, \bibinfo {author} {\bibfnamefont {H.~F.}\ \bibnamefont
  {Robey}}, \bibinfo {author} {\bibfnamefont {B.~K.}\ \bibnamefont {Spears}},
  \bibinfo {author} {\bibfnamefont {P.~T.}\ \bibnamefont {Springer}}, \bibinfo
  {author} {\bibfnamefont {L.~J.}\ \bibnamefont {Suter}}, \bibinfo {author}
  {\bibfnamefont {C.~A.}\ \bibnamefont {Thomas}}, \bibinfo {author}
  {\bibfnamefont {R.~P.}\ \bibnamefont {Town}}, \bibinfo {author}
  {\bibfnamefont {R.}~\bibnamefont {Vesey}}, \bibinfo {author} {\bibfnamefont
  {S.~V.}\ \bibnamefont {Weber}}, \bibinfo {author} {\bibfnamefont {H.~L.}\
  \bibnamefont {Wilkens}}, \ and\ \bibinfo {author} {\bibfnamefont {D.~C.}\
  \bibnamefont {Wilson}},\ }\href {\doibase
  http://dx.doi.org/10.1063/1.3592169} {\bibfield  {journal} {\bibinfo
  {journal} {Physics of Plasmas (1994-present)}\ }\textbf {\bibinfo {volume}
  {18}},\ \bibinfo {eid} {051001} (\bibinfo {year} {2011})}\BibitemShut
  {NoStop}%
\bibitem [{\citenamefont {Benedict}\ \emph {et~al.}(2009)\citenamefont
  {Benedict}, \citenamefont {Ogitsu}, \citenamefont {Trave}, \citenamefont
  {Wu}, \citenamefont {Sterne},\ and\ \citenamefont {Schwegler}}]{benedict-be}%
  \BibitemOpen
  \bibfield  {author} {\bibinfo {author} {\bibfnamefont {L.~X.}\ \bibnamefont
  {Benedict}}, \bibinfo {author} {\bibfnamefont {T.}~\bibnamefont {Ogitsu}},
  \bibinfo {author} {\bibfnamefont {A.}~\bibnamefont {Trave}}, \bibinfo
  {author} {\bibfnamefont {C.~J.}\ \bibnamefont {Wu}}, \bibinfo {author}
  {\bibfnamefont {P.~A.}\ \bibnamefont {Sterne}}, \ and\ \bibinfo {author}
  {\bibfnamefont {E.}~\bibnamefont {Schwegler}},\ }\href {\doibase
  10.1103/PhysRevB.79.064106} {\bibfield  {journal} {\bibinfo  {journal} {Phys.
  Rev. B}\ }\textbf {\bibinfo {volume} {79}},\ \bibinfo {pages} {064106}
  (\bibinfo {year} {2009})}\BibitemShut {NoStop}%
\bibitem [{\citenamefont {Lazicki}\ \emph {et~al.}(2012)\citenamefont
  {Lazicki}, \citenamefont {Dewaele}, \citenamefont {Loubeyre},\ and\
  \citenamefont {Mezouar}}]{lazicki-be}%
  \BibitemOpen
  \bibfield  {author} {\bibinfo {author} {\bibfnamefont {A.}~\bibnamefont
  {Lazicki}}, \bibinfo {author} {\bibfnamefont {A.}~\bibnamefont {Dewaele}},
  \bibinfo {author} {\bibfnamefont {P.}~\bibnamefont {Loubeyre}}, \ and\
  \bibinfo {author} {\bibfnamefont {M.}~\bibnamefont {Mezouar}},\ }\href
  {\doibase 10.1103/PhysRevB.86.174118} {\bibfield  {journal} {\bibinfo
  {journal} {Phys. Rev. B}\ }\textbf {\bibinfo {volume} {86}},\ \bibinfo
  {pages} {174118} (\bibinfo {year} {2012})}\BibitemShut {NoStop}%
\bibitem [{\citenamefont {Zhang}\ \emph {et~al.}(2013)\citenamefont {Zhang},
  \citenamefont {Zhu}, \citenamefont {Velisavljevic}, \citenamefont {Wang},\
  and\ \citenamefont {Zhao}}]{zhang-dac}%
  \BibitemOpen
  \bibfield  {author} {\bibinfo {author} {\bibfnamefont {J.}~\bibnamefont
  {Zhang}}, \bibinfo {author} {\bibfnamefont {J.}~\bibnamefont {Zhu}}, \bibinfo
  {author} {\bibfnamefont {N.}~\bibnamefont {Velisavljevic}}, \bibinfo {author}
  {\bibfnamefont {L.}~\bibnamefont {Wang}}, \ and\ \bibinfo {author}
  {\bibfnamefont {Y.}~\bibnamefont {Zhao}},\ }\href {\doibase
  http://dx.doi.org/10.1063/1.4828886} {\bibfield  {journal} {\bibinfo
  {journal} {Journal of Applied Physics}\ }\textbf {\bibinfo {volume} {114}},\
  \bibinfo {eid} {173509} (\bibinfo {year} {2013})}\BibitemShut {NoStop}%
\bibitem [{\citenamefont {Evans}\ \emph {et~al.}(2005)\citenamefont {Evans},
  \citenamefont {Lipp}, \citenamefont {Cynn}, \citenamefont {Yoo},
  \citenamefont {Somayazulu}, \citenamefont {H\"ausermann}, \citenamefont
  {Shen},\ and\ \citenamefont {Prakapenka}}]{evans-dac}%
  \BibitemOpen
  \bibfield  {author} {\bibinfo {author} {\bibfnamefont {W.~J.}\ \bibnamefont
  {Evans}}, \bibinfo {author} {\bibfnamefont {M.~J.}\ \bibnamefont {Lipp}},
  \bibinfo {author} {\bibfnamefont {H.}~\bibnamefont {Cynn}}, \bibinfo {author}
  {\bibfnamefont {C.~S.}\ \bibnamefont {Yoo}}, \bibinfo {author} {\bibfnamefont
  {M.}~\bibnamefont {Somayazulu}}, \bibinfo {author} {\bibfnamefont
  {D.}~\bibnamefont {H\"ausermann}}, \bibinfo {author} {\bibfnamefont
  {G.}~\bibnamefont {Shen}}, \ and\ \bibinfo {author} {\bibfnamefont
  {V.}~\bibnamefont {Prakapenka}},\ }\href {\doibase
  10.1103/PhysRevB.72.094113} {\bibfield  {journal} {\bibinfo  {journal} {Phys.
  Rev. B}\ }\textbf {\bibinfo {volume} {72}},\ \bibinfo {pages} {094113}
  (\bibinfo {year} {2005})}\BibitemShut {NoStop}%
\bibitem [{\citenamefont {Brown}\ \emph {et~al.}(2014)\citenamefont {Brown},
  \citenamefont {Knudson}, \citenamefont {Alexander},\ and\ \citenamefont
  {Asay}}]{brown-be-dynamic}%
  \BibitemOpen
  \bibfield  {author} {\bibinfo {author} {\bibfnamefont {J.~L.}\ \bibnamefont
  {Brown}}, \bibinfo {author} {\bibfnamefont {M.~D.}\ \bibnamefont {Knudson}},
  \bibinfo {author} {\bibfnamefont {C.~S.}\ \bibnamefont {Alexander}}, \ and\
  \bibinfo {author} {\bibfnamefont {J.~R.}\ \bibnamefont {Asay}},\ }\href
  {\doibase http://dx.doi.org/10.1063/1.4890232} {\bibfield  {journal}
  {\bibinfo  {journal} {Journal of Applied Physics}\ }\textbf {\bibinfo
  {volume} {116}},\ \bibinfo {eid} {033502} (\bibinfo {year}
  {2014})}\BibitemShut {NoStop}%
\bibitem [{\citenamefont {Robert}\ and\ \citenamefont
  {Sollier}(2006)}]{french-be-calc}%
  \BibitemOpen
  \bibfield  {author} {\bibinfo {author} {\bibfnamefont {G.}~\bibnamefont
  {Robert}}\ and\ \bibinfo {author} {\bibfnamefont {A.}~\bibnamefont
  {Sollier}},\ }\href@noop {} {\bibfield  {journal} {\bibinfo  {journal}
  {Journal de Physique IV (Proceedings)}\ }\textbf {\bibinfo {volume} {134}},\
  \bibinfo {pages} {257} (\bibinfo {year} {2006})}\BibitemShut {NoStop}%
\bibitem [{\citenamefont {Robert}\ \emph {et~al.}(2010)\citenamefont {Robert},
  \citenamefont {Legrand},\ and\ \citenamefont {Bernard}}]{robert-be-calc}%
  \BibitemOpen
  \bibfield  {author} {\bibinfo {author} {\bibfnamefont {G.}~\bibnamefont
  {Robert}}, \bibinfo {author} {\bibfnamefont {P.}~\bibnamefont {Legrand}}, \
  and\ \bibinfo {author} {\bibfnamefont {S.}~\bibnamefont {Bernard}},\ }\href
  {\doibase 10.1103/PhysRevB.82.104118} {\bibfield  {journal} {\bibinfo
  {journal} {Phys. Rev. B}\ }\textbf {\bibinfo {volume} {82}},\ \bibinfo
  {pages} {104118} (\bibinfo {year} {2010})}\BibitemShut {NoStop}%
\bibitem [{\citenamefont {Luo}\ \emph {et~al.}(2012)\citenamefont {Luo},
  \citenamefont {Cai}, \citenamefont {Chen}, \citenamefont {Jing},\ and\
  \citenamefont {Alf{\`e}}}]{luo-alfe-be-calc}%
  \BibitemOpen
  \bibfield  {author} {\bibinfo {author} {\bibfnamefont {F.}~\bibnamefont
  {Luo}}, \bibinfo {author} {\bibfnamefont {L.-C.}\ \bibnamefont {Cai}},
  \bibinfo {author} {\bibfnamefont {X.-R.}\ \bibnamefont {Chen}}, \bibinfo
  {author} {\bibfnamefont {F.-Q.}\ \bibnamefont {Jing}}, \ and\ \bibinfo
  {author} {\bibfnamefont {D.}~\bibnamefont {Alf{\`e}}},\ }\href {\doibase
  http://dx.doi.org/10.1063/1.3688344} {\bibfield  {journal} {\bibinfo
  {journal} {Journal of Applied Physics}\ }\textbf {\bibinfo {volume} {111}},\
  \bibinfo {eid} {053503} (\bibinfo {year} {2012})}\BibitemShut {NoStop}%
\bibitem [{\citenamefont {Cheng}\ \emph {et~al.}(2013)\citenamefont {Cheng},
  \citenamefont {Chen}, \citenamefont {Xue}, \citenamefont {Ji},\ and\
  \citenamefont {Gong}}]{cheng-be-calc}%
  \BibitemOpen
  \bibfield  {author} {\bibinfo {author} {\bibfnamefont {Y.}~\bibnamefont
  {Cheng}}, \bibinfo {author} {\bibfnamefont {H.-H.}\ \bibnamefont {Chen}},
  \bibinfo {author} {\bibfnamefont {F.-X.}\ \bibnamefont {Xue}}, \bibinfo
  {author} {\bibfnamefont {G.-F.}\ \bibnamefont {Ji}}, \ and\ \bibinfo {author}
  {\bibfnamefont {M.}~\bibnamefont {Gong}},\ }\href {\doibase
  10.1142/S0217979213501300} {\bibfield  {journal} {\bibinfo  {journal}
  {International Journal of Modern Physics B}\ }\textbf {\bibinfo {volume}
  {27}},\ \bibinfo {pages} {1350130} (\bibinfo {year} {2013})}\BibitemShut
  {NoStop}%
\bibitem [{\citenamefont {Guo}\ \emph {et~al.}(2014)\citenamefont {Guo},
  \citenamefont {Luo},\ and\ \citenamefont {Cheng}}]{guo-be-calc}%
  \BibitemOpen
  \bibfield  {author} {\bibinfo {author} {\bibfnamefont {Z.-C.}\ \bibnamefont
  {Guo}}, \bibinfo {author} {\bibfnamefont {F.}~\bibnamefont {Luo}}, \ and\
  \bibinfo {author} {\bibfnamefont {Y.}~\bibnamefont {Cheng}},\ }\href
  {\doibase http://dx.doi.org/10.1016/j.commatsci.2013.12.013} {\bibfield
  {journal} {\bibinfo  {journal} {Computational Materials Science}\ }\textbf
  {\bibinfo {volume} {84}},\ \bibinfo {pages} {139 } (\bibinfo {year}
  {2014})}\BibitemShut {NoStop}%
\bibitem [{\citenamefont {Perdew}\ and\ \citenamefont
  {Schmidt}(2001)}]{jacobs-ladder}%
  \BibitemOpen
  \bibfield  {author} {\bibinfo {author} {\bibfnamefont {J.~P.}\ \bibnamefont
  {Perdew}}\ and\ \bibinfo {author} {\bibfnamefont {K.}~\bibnamefont
  {Schmidt}},\ }in\ \href@noop {} {\emph {\bibinfo {booktitle} {AIP Conference
  Proceedings}}}\ (\bibinfo {organization} {IOP INSTITUTE OF PHYSICS PUBLISHING
  LTD},\ \bibinfo {year} {2001})\ pp.\ \bibinfo {pages} {1--20}\BibitemShut
  {NoStop}%
\bibitem [{\citenamefont {Kim}\ \emph {et~al.}(2012)\citenamefont {Kim},
  \citenamefont {Esler}, \citenamefont {McMinis}, \citenamefont {Morales},
  \citenamefont {Clark}, \citenamefont {Shulenburger},\ and\ \citenamefont
  {Ceperley}}]{qmcpack}%
  \BibitemOpen
  \bibfield  {author} {\bibinfo {author} {\bibfnamefont {J.}~\bibnamefont
  {Kim}}, \bibinfo {author} {\bibfnamefont {K.~P.}\ \bibnamefont {Esler}},
  \bibinfo {author} {\bibfnamefont {J.}~\bibnamefont {McMinis}}, \bibinfo
  {author} {\bibfnamefont {M.~A.}\ \bibnamefont {Morales}}, \bibinfo {author}
  {\bibfnamefont {B.~K.}\ \bibnamefont {Clark}}, \bibinfo {author}
  {\bibfnamefont {L.}~\bibnamefont {Shulenburger}}, \ and\ \bibinfo {author}
  {\bibfnamefont {D.~M.}\ \bibnamefont {Ceperley}},\ }\href@noop {} {\bibfield
  {journal} {\bibinfo  {journal} {Journal of Physics: Conference Series}\
  }\textbf {\bibinfo {volume} {402}},\ \bibinfo {pages} {012008} (\bibinfo
  {year} {2012})}\BibitemShut {NoStop}%
\bibitem [{\citenamefont {Esler}\ \emph {et~al.}(2012)\citenamefont {Esler},
  \citenamefont {Kim}, \citenamefont {Shulenburger},\ and\ \citenamefont
  {Ceperley}}]{gpu-qmc}%
  \BibitemOpen
  \bibfield  {author} {\bibinfo {author} {\bibfnamefont {K.~P.}~\bibnamefont
  {Esler}}, \bibinfo {author} {\bibfnamefont {J.}~\bibnamefont {Kim}}, \bibinfo
  {author} {\bibfnamefont {L.}~\bibnamefont {Shulenburger}}, \ and\ \bibinfo
  {author} {\bibfnamefont {D.}~\bibnamefont {Ceperley}},\ }\href@noop {}
  {\bibfield  {journal} {\bibinfo  {journal} {Computing in Science and
  Engineering}\ }\textbf {\bibinfo {volume} {14}},\ \bibinfo {pages} {40}
  (\bibinfo {year} {2012})}\BibitemShut {NoStop}%
\bibitem [{opi()}]{opium}%
  \BibitemOpen
  \href@noop {} {\enquote {\bibinfo {title} {Opium pseudopotential package},}\
  }\BibitemShut {NoStop}%
\bibitem [{\citenamefont {Troullier}\ and\ \citenamefont
  {Martins}(1991)}]{troullier-martins}%
  \BibitemOpen
  \bibfield  {author} {\bibinfo {author} {\bibfnamefont {N.}~\bibnamefont
  {Troullier}}\ and\ \bibinfo {author} {\bibfnamefont {J.~L.}\ \bibnamefont
  {Martins}},\ }\href {\doibase 10.1103/PhysRevB.43.1993} {\bibfield  {journal}
  {\bibinfo  {journal} {Phys. Rev. B}\ }\textbf {\bibinfo {volume} {43}},\
  \bibinfo {pages} {1993} (\bibinfo {year} {1991})}\BibitemShut {NoStop}%
\bibitem [{\citenamefont {Casula}(2006)}]{casula-t-moves}%
  \BibitemOpen
  \bibfield  {author} {\bibinfo {author} {\bibfnamefont {M.}~\bibnamefont
  {Casula}},\ }\href@noop {} {\bibfield  {journal} {\bibinfo  {journal}
  {Physical Review B}\ }\textbf {\bibinfo {volume} {74}},\ \bibinfo {pages}
  {161102} (\bibinfo {year} {2006})}\BibitemShut {NoStop}%
\bibitem [{\citenamefont {Fraser}\ \emph {et~al.}(1996)\citenamefont {Fraser},
  \citenamefont {Foulkes}, \citenamefont {Rajagopal}, \citenamefont {Needs},
  \citenamefont {Kenny},\ and\ \citenamefont {Williamson}}]{MPC}%
  \BibitemOpen
  \bibfield  {author} {\bibinfo {author} {\bibfnamefont {L.~M.}\ \bibnamefont
  {Fraser}}, \bibinfo {author} {\bibfnamefont {W.~M.~C.}\ \bibnamefont
  {Foulkes}}, \bibinfo {author} {\bibfnamefont {G.}~\bibnamefont {Rajagopal}},
  \bibinfo {author} {\bibfnamefont {R.~J.}\ \bibnamefont {Needs}}, \bibinfo
  {author} {\bibfnamefont {S.~D.}\ \bibnamefont {Kenny}}, \ and\ \bibinfo
  {author} {\bibfnamefont {A.~J.}\ \bibnamefont {Williamson}},\ }\href
  {\doibase 10.1103/PhysRevB.53.1814} {\bibfield  {journal} {\bibinfo
  {journal} {Phys. Rev. B}\ }\textbf {\bibinfo {volume} {53}},\ \bibinfo
  {pages} {1814} (\bibinfo {year} {1996})}\BibitemShut {NoStop}%
\bibitem [{\citenamefont {Chiesa}\ \emph {et~al.}(2006)\citenamefont {Chiesa},
  \citenamefont {Ceperley}, \citenamefont {Martin},\ and\ \citenamefont
  {Holzmann}}]{Chiesa-finite-size}%
  \BibitemOpen
  \bibfield  {author} {\bibinfo {author} {\bibfnamefont {S.}~\bibnamefont
  {Chiesa}}, \bibinfo {author} {\bibfnamefont {D.~M.}\ \bibnamefont
  {Ceperley}}, \bibinfo {author} {\bibfnamefont {R.~M.}\ \bibnamefont
  {Martin}}, \ and\ \bibinfo {author} {\bibfnamefont {M.}~\bibnamefont
  {Holzmann}},\ }\href {\doibase 10.1103/PhysRevLett.97.076404} {\bibfield
  {journal} {\bibinfo  {journal} {Phys. Rev. Lett.}\ }\textbf {\bibinfo
  {volume} {97}},\ \bibinfo {pages} {076404} (\bibinfo {year}
  {2006})}\BibitemShut {NoStop}%
\bibitem [{\citenamefont {Mackay}\ and\ \citenamefont {Hill}(1963)}]{expt-c-a}%
  \BibitemOpen
  \bibfield  {author} {\bibinfo {author} {\bibfnamefont {K.}~\bibnamefont
  {Mackay}}\ and\ \bibinfo {author} {\bibfnamefont {N.}~\bibnamefont {Hill}},\
  }\href@noop {} {\bibfield  {journal} {\bibinfo  {journal} {Journal of Nuclear
  Materials}\ }\textbf {\bibinfo {volume} {8}},\ \bibinfo {pages} {263}
  (\bibinfo {year} {1963})}\BibitemShut {NoStop}%
\bibitem [{\citenamefont {Vinet}\ \emph {et~al.}(1986)\citenamefont {Vinet},
  \citenamefont {Ferrante}, \citenamefont {Smith},\ and\ \citenamefont
  {Rose}}]{vinet}%
  \BibitemOpen
  \bibfield  {author} {\bibinfo {author} {\bibfnamefont {P.}~\bibnamefont
  {Vinet}}, \bibinfo {author} {\bibfnamefont {J.}~\bibnamefont {Ferrante}},
  \bibinfo {author} {\bibfnamefont {J.~R.}\ \bibnamefont {Smith}}, \ and\
  \bibinfo {author} {\bibfnamefont {J.~H.}\ \bibnamefont {Rose}},\ }\href
  {http://stacks.iop.org/0022-3719/19/i=20/a=001} {\bibfield  {journal}
  {\bibinfo  {journal} {Journal of Physics C: Solid State Physics}\ }\textbf
  {\bibinfo {volume} {19}},\ \bibinfo {pages} {L467} (\bibinfo {year}
  {1986})}\BibitemShut {NoStop}%
\bibitem [{\citenamefont {Esler}\ \emph {et~al.}(2010)\citenamefont {Esler},
  \citenamefont {Cohen}, \citenamefont {Militzer}, \citenamefont {Kim},
  \citenamefont {Needs},\ and\ \citenamefont {Towler}}]{esler-cBN}%
  \BibitemOpen
  \bibfield  {author} {\bibinfo {author} {\bibfnamefont {K.~P.}\ \bibnamefont
  {Esler}}, \bibinfo {author} {\bibfnamefont {R.~E.}\ \bibnamefont {Cohen}},
  \bibinfo {author} {\bibfnamefont {B.}~\bibnamefont {Militzer}}, \bibinfo
  {author} {\bibfnamefont {J.}~\bibnamefont {Kim}}, \bibinfo {author}
  {\bibfnamefont {R.~J.}\ \bibnamefont {Needs}}, \ and\ \bibinfo {author}
  {\bibfnamefont {M.~D.}\ \bibnamefont {Towler}},\ }\href@noop {} {\bibfield
  {journal} {\bibinfo  {journal} {Physical Review Letters}\ }\textbf {\bibinfo
  {volume} {104}},\ \bibinfo {pages} {185702} (\bibinfo {year}
  {2010})}\BibitemShut {NoStop}%
\bibitem [{\citenamefont {Giannozzi}\ \emph {et~al.}(2009)\citenamefont
  {Giannozzi}, \citenamefont {Baroni}, \citenamefont {Bonini}, \citenamefont
  {Calandra}, \citenamefont {Car}, \citenamefont {Cavazzoni}, \citenamefont
  {Ceresoli}, \citenamefont {Chiarotti}, \citenamefont {Cococcioni},
  \citenamefont {Dabo}, \citenamefont {{Dal Corso}}, \citenamefont
  {de~Gironcoli}, \citenamefont {Fabris}, \citenamefont {Fratesi},
  \citenamefont {Gebauer}, \citenamefont {Gerstmann}, \citenamefont
  {Gougoussis}, \citenamefont {Kokalj}, \citenamefont {Lazzeri}, \citenamefont
  {Martin-Samos}, \citenamefont {Marzari}, \citenamefont {Mauri}, \citenamefont
  {Mazzarello}, \citenamefont {Paolini}, \citenamefont {Pasquarello},
  \citenamefont {Paulatto}, \citenamefont {Sbraccia}, \citenamefont {Scandolo},
  \citenamefont {Sclauzero}, \citenamefont {Seitsonen}, \citenamefont
  {Smogunov}, \citenamefont {Umari},\ and\ \citenamefont
  {Wentzcovitch}}]{quantum-espresso}%
  \BibitemOpen
  \bibfield  {author} {\bibinfo {author} {\bibfnamefont {P.}~\bibnamefont
  {Giannozzi}}, \bibinfo {author} {\bibfnamefont {S.}~\bibnamefont {Baroni}},
  \bibinfo {author} {\bibfnamefont {N.}~\bibnamefont {Bonini}}, \bibinfo
  {author} {\bibfnamefont {M.}~\bibnamefont {Calandra}}, \bibinfo {author}
  {\bibfnamefont {R.}~\bibnamefont {Car}}, \bibinfo {author} {\bibfnamefont
  {C.}~\bibnamefont {Cavazzoni}}, \bibinfo {author} {\bibfnamefont
  {D.}~\bibnamefont {Ceresoli}}, \bibinfo {author} {\bibfnamefont {G.~L.}\
  \bibnamefont {Chiarotti}}, \bibinfo {author} {\bibfnamefont {M.}~\bibnamefont
  {Cococcioni}}, \bibinfo {author} {\bibfnamefont {I.}~\bibnamefont {Dabo}},
  \bibinfo {author} {\bibfnamefont {A.}~\bibnamefont {{Dal Corso}}}, \bibinfo
  {author} {\bibfnamefont {S.}~\bibnamefont {de~Gironcoli}}, \bibinfo {author}
  {\bibfnamefont {S.}~\bibnamefont {Fabris}}, \bibinfo {author} {\bibfnamefont
  {G.}~\bibnamefont {Fratesi}}, \bibinfo {author} {\bibfnamefont
  {R.}~\bibnamefont {Gebauer}}, \bibinfo {author} {\bibfnamefont
  {U.}~\bibnamefont {Gerstmann}}, \bibinfo {author} {\bibfnamefont
  {C.}~\bibnamefont {Gougoussis}}, \bibinfo {author} {\bibfnamefont
  {A.}~\bibnamefont {Kokalj}}, \bibinfo {author} {\bibfnamefont
  {M.}~\bibnamefont {Lazzeri}}, \bibinfo {author} {\bibfnamefont
  {L.}~\bibnamefont {Martin-Samos}}, \bibinfo {author} {\bibfnamefont
  {N.}~\bibnamefont {Marzari}}, \bibinfo {author} {\bibfnamefont
  {F.}~\bibnamefont {Mauri}}, \bibinfo {author} {\bibfnamefont
  {R.}~\bibnamefont {Mazzarello}}, \bibinfo {author} {\bibfnamefont
  {S.}~\bibnamefont {Paolini}}, \bibinfo {author} {\bibfnamefont
  {A.}~\bibnamefont {Pasquarello}}, \bibinfo {author} {\bibfnamefont
  {L.}~\bibnamefont {Paulatto}}, \bibinfo {author} {\bibfnamefont
  {C.}~\bibnamefont {Sbraccia}}, \bibinfo {author} {\bibfnamefont
  {S.}~\bibnamefont {Scandolo}}, \bibinfo {author} {\bibfnamefont
  {G.}~\bibnamefont {Sclauzero}}, \bibinfo {author} {\bibfnamefont {A.~P.}\
  \bibnamefont {Seitsonen}}, \bibinfo {author} {\bibfnamefont {A.}~\bibnamefont
  {Smogunov}}, \bibinfo {author} {\bibfnamefont {P.}~\bibnamefont {Umari}}, \
  and\ \bibinfo {author} {\bibfnamefont {R.~M.}\ \bibnamefont {Wentzcovitch}},\
  }\href {http://www.quantum-espresso.org} {\bibfield  {journal} {\bibinfo
  {journal} {Journal of Physics: Condensed Matter}\ }\textbf {\bibinfo {volume}
  {21}},\ \bibinfo {pages} {395502 (19pp)} (\bibinfo {year}
  {2009})}\BibitemShut {NoStop}%
\bibitem [{\citenamefont {Ceperley}(1986)}]{ceperley1986statistical}%
  \BibitemOpen
  \bibfield  {author} {\bibinfo {author} {\bibfnamefont {D.}~\bibnamefont
  {Ceperley}},\ }\href@noop {} {\bibfield  {journal} {\bibinfo  {journal}
  {Journal of Statistical Physics}\ }\textbf {\bibinfo {volume} {43}},\
  \bibinfo {pages} {815} (\bibinfo {year} {1986})}\BibitemShut {NoStop}%
\bibitem [{\citenamefont {Ma}\ \emph {et~al.}(2005)\citenamefont {Ma},
  \citenamefont {Drummond}, \citenamefont {Towler},\ and\ \citenamefont
  {Needs}}]{ma2005all}%
  \BibitemOpen
  \bibfield  {author} {\bibinfo {author} {\bibfnamefont {A.}~\bibnamefont
  {Ma}}, \bibinfo {author} {\bibfnamefont {N.~D.}\ \bibnamefont {Drummond}},
  \bibinfo {author} {\bibfnamefont {M.~D.}\ \bibnamefont {Towler}}, \ and\ \bibinfo
  {author} {\bibfnamefont {R.~J.}\ \bibnamefont {Needs}},\ }\href@noop {}
  {\bibfield  {journal} {\bibinfo  {journal} {Physical Review E}\ }\textbf
  {\bibinfo {volume} {71}},\ \bibinfo {pages} {066704} (\bibinfo {year}
  {2005})}\BibitemShut {NoStop}%
\bibitem [{\citenamefont {Trail}\ and\ \citenamefont
  {Needs}(2005)}]{Trail-Needs-PPs}%
  \BibitemOpen
  \bibfield  {author} {\bibinfo {author} {\bibfnamefont {J.~R.}\ \bibnamefont
  {Trail}}\ and\ \bibinfo {author} {\bibfnamefont {R.~J.}\ \bibnamefont
  {Needs}},\ }\href {\doibase http://dx.doi.org/10.1063/1.1888569} {\bibfield
  {journal} {\bibinfo  {journal} {The Journal of Chemical Physics}\ }\textbf
  {\bibinfo {volume} {122}},\ \bibinfo {eid} {174109} (\bibinfo {year}
  {2005})}\BibitemShut {NoStop}%
\bibitem [{\citenamefont {Burkatzki}\ \emph {et~al.}(2007)\citenamefont
  {Burkatzki}, \citenamefont {Filippi},\ and\ \citenamefont
  {Dolg}}]{burkatzki-filippi-dolg-pps}%
  \BibitemOpen
  \bibfield  {author} {\bibinfo {author} {\bibfnamefont {M.}~\bibnamefont
  {Burkatzki}}, \bibinfo {author} {\bibfnamefont {C.}~\bibnamefont {Filippi}},
  \ and\ \bibinfo {author} {\bibfnamefont {M.}~\bibnamefont {Dolg}},\
  }\href@noop {} {\bibfield  {journal} {\bibinfo  {journal} {The Journal of
  chemical physics}\ }\textbf {\bibinfo {volume} {126}},\ \bibinfo {pages}
  {234105} (\bibinfo {year} {2007})}\BibitemShut {NoStop}%
\bibitem [{\citenamefont {Trail}\ and\ \citenamefont
  {Needs}(2013)}]{Trail-Needs-CEPPs}%
  \BibitemOpen
  \bibfield  {author} {\bibinfo {author} {\bibfnamefont {J.~R.}\ \bibnamefont
  {Trail}}\ and\ \bibinfo {author} {\bibfnamefont {R.~J.}\ \bibnamefont
  {Needs}},\ }\href {\doibase http://dx.doi.org/10.1063/1.4811651} {\bibfield
  {journal} {\bibinfo  {journal} {The Journal of Chemical Physics}\ }\textbf
  {\bibinfo {volume} {139}},\ \bibinfo {eid} {014101} (\bibinfo {year}
  {2013})}\BibitemShut {NoStop}%
\bibitem [{\citenamefont {Kleinman}\ and\ \citenamefont
  {Bylander}(1982)}]{kleinmann-bylander}%
  \BibitemOpen
  \bibfield  {author} {\bibinfo {author} {\bibfnamefont {L.}~\bibnamefont
  {Kleinman}}\ and\ \bibinfo {author} {\bibfnamefont {D.~M.}\ \bibnamefont
  {Bylander}},\ }\href@noop {} {\bibfield  {journal} {\bibinfo  {journal}
  {Physical Review Letters}\ }\textbf {\bibinfo {volume} {48}},\ \bibinfo
  {pages} {1425} (\bibinfo {year} {1982})}\BibitemShut {NoStop}%
\end{thebibliography}%

%\begin{thebibliography}{10}

%\end{thebibliography}

\end{document}